\documentclass[conference,  letterpaper]{IEEEtran}

\usepackage{graphicx} 
\usepackage{subcaption}
\usepackage[utf8]{inputenc}
\usepackage{cite}
\usepackage[cmex10]{amsmath}
\usepackage{caption}
\usepackage{amsthm}
\usepackage{xcolor}

    {
    \theoremstyle{plain}
    \newtheorem{theorem}{Theorem}

\newtheorem{lemma}{Lemma}

\newtheorem{definition}{Definition}

    }
\usepackage{thm-restate}
\usepackage{comment}
\usepackage{amssymb}
\usepackage{mathtools}
\usepackage{dsfont}

\DeclareMathOperator{\sgn}{sgn}
\graphicspath{{Images-Graphs/}}

\def \Ebb {\mathbb{E}}

\def \Ecal {\mathcal{E}}

\title{Adversarial Quantum Machine Learning: An Information-Theoretic Generalization Analysis} 
\author{
  \IEEEauthorblockN{Petros Georgiou, Sharu Theresa Jose}
  \IEEEauthorblockA{Department of Computer Science,
               \\University of Birmingham, UK\\
               Email: pxg402@student.bham.ac.uk, s.t.jose@bham.ac.uk}
  \and
  \IEEEauthorblockN{Osvaldo Simeone}
  \IEEEauthorblockA{KCLIP lab\\ Centre for Intelligent Information Processing Systems (CIIPS)
\\
               Department of Engineering, King’s College London\\
               Email: osvaldo.simeone@kcl.ac.uk}
   
}

\begin{document}
\maketitle

\begin{abstract}
   In a manner analogous to their classical counterparts, quantum classifiers are vulnerable to adversarial attacks that perturb their inputs. A promising countermeasure is to train the quantum classifier by adopting an attack-aware, or adversarial, loss function. This paper studies the generalization properties of  quantum classifiers that are  adversarially trained   against bounded-norm white-box attacks. Specifically, a quantum adversary maximizes the classifier's loss by transforming an input state $\rho(x)$ into a state $\lambda$ that is $\epsilon$-close to the original state $\rho(x)$ in $p$-Schatten distance.
    Under suitable assumptions on the quantum embedding $\rho(x)$, we derive novel information-theoretic upper bounds on the generalization error of adversarially trained quantum classifiers  for $p=1$ and $p=\infty$. The derived upper bounds consist of two terms: the first is an exponential function of the  2-R{\'e}nyi mutual information between classical data and quantum embedding, while the second term scales linearly with the  adversarial perturbation size $\epsilon$. Both terms 
    are shown to decrease as $1/\sqrt{T}$ over the training set size $T$. 
    An extension is also considered in which the adversary assumed during training has  different parameters $p$ and $\epsilon$ as compared to the adversary affecting the test inputs. 
    Finally, we validate our theoretical findings with numerical experiments for a synthetic setting.
\end{abstract}

\section{Introduction}

\noindent \emph{Motivation}: Quantum machine learning (QML) has emerged as a design paradigm for current noisy intermediate scale quantum (NISQ) computers \cite{schuld2021machine,simeone2022introduction}. Among the main projected application of QML is data analytics, of  which classification is a prototypical example.  As shown in Fig. 1(a), in a typical quantum classification problem, a classical input $x$ -- such as an image, a text, or a vector of tunable parameters for a physical experiment -- is mapped to a quantum state $\rho(x)$, which is known as a \emph{quantum embedding}. The quantum embedding map $\rho(x)$ may be implemented by a  quantum circuit or by some physical mechanism, possibly encompassing also quantum sensing \cite{davidovich2024quantum}. The design goal  is to find a classifier, consisting of a positive operator valued measure (POVM), that can predict the true class $c$ associated with input $x$ with reasonable accuracy. 

\begin{figure}
    \centering
    \includegraphics[scale=0.28, clip=true, trim=0.35in 2.2in 0.3in 1.2in]{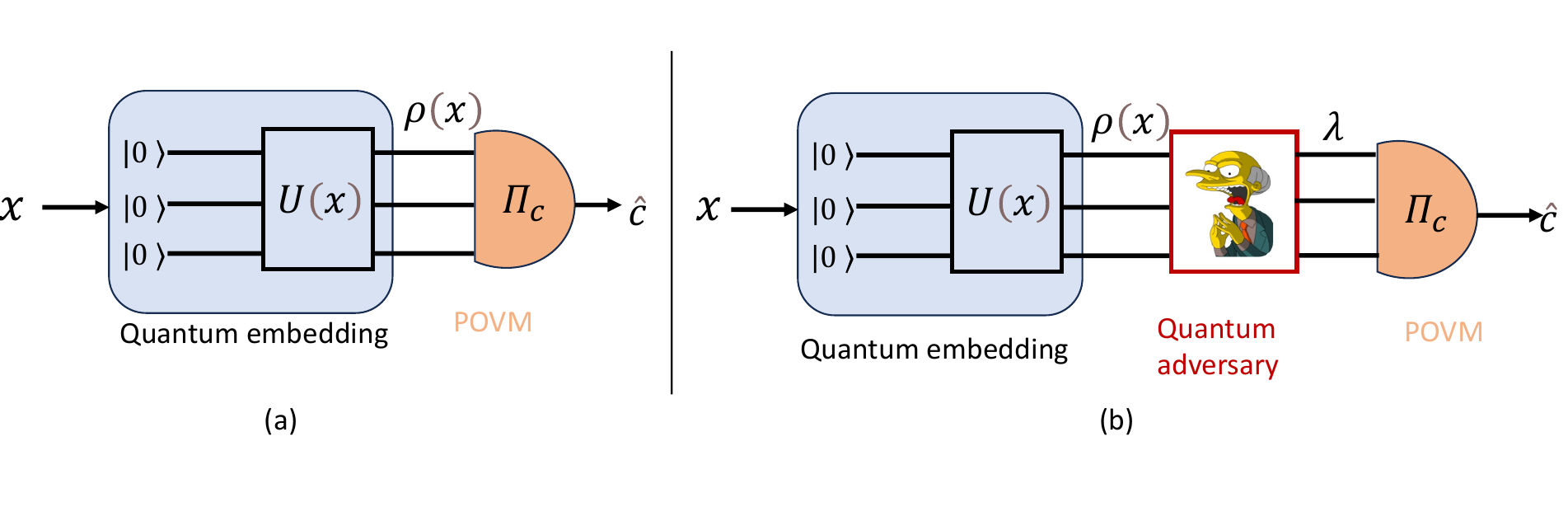}
    \caption{Quantum classification in $(a)$ a non-adversarial setting, in which the quantum measurement $\Pi$ acts on the unperturbed quantum embedding $\rho(x)$; $(b)$ an adversarial setting, in which  the state $\rho(x)$ is perturbed by a quantum adversary to yield a state $\lambda$.}
    \label{fig:1}
    \vspace{-0.4cm}
\end{figure}

Despite quantum classifiers having shown promising results \cite{havlivcek2019supervised}, recent works \cite{Lu_2020, west2023benchmarking,ren2022experimental}  have highlighted their vulnerability to adversarial attacks. A \textit{quantum adversary} can perturb the input quantum state $\rho(x)$ via the application of a quantum channel, producing a state $\lambda$ for which the classifier is less likely to identify the true class $c$. 



\textit{Adversarial training} was found to be a promising defense strategy \cite{Lu_2020,ren2022experimental}. In adversarial training, the classifier replaces the conventional classification loss with an \textit{adversarial loss}  that accounts for the worst-case effect of an  adversarial perturbation of the quantum embedding. This approach results in a min-max optimization problem with outer minimization over POVMs and inner maximization over adversarial perturbations. Our aim is to understand how well an adversarially trained classifier \textit{generalizes} to new, previously unseen quantum states subjected to a possibly different adversarial attack. 


\noindent \emph{Related Work}:  While the theory of adversarial generalization has recently garnered attention in classical adversarial machine learning \cite{yin2019rademacher, awasthi2020adversarial,xiao2022adversarial}, related efforts have not been reported for QML. Indeed, existing works on the generalization analysis of QML models focus on the conventional non-adversarial setting \cite{banchi2021generalization,caro2022generalization,caro2023out}. Our work is particularly inspired by \cite{banchi2021generalization}, which presented an information-theoretic analysis of generalization for quantum classifier in the absence of quantum adversaries. Our generalization bounds extend those derived in \cite{banchi2021generalization} by accounting for the impact of adversarial training and for the presence of a quantum attacker at test time. 


\noindent \emph{Main Contributions}: In this work, we study quantum adversarial attacks which perturb the input quantum state $\rho(x)$ to a state that is $\epsilon$-close to $\rho(x)$ in  $p$-Schatten distance. Our main contributions are as follows:\\
\noindent $\bullet$ We derive  new information-theoretic upper bounds on the adversarial generalization error for $p=1$ and $p=\infty$. The resulting upper bounds consist of two terms: The first, which coincides with the bound in \cite{banchi2021generalization}, captures the non-adversarial generalization error  via the exponentiated 2-R{\'e}nyi-mutual information  between the classical input and the quantum embedding; while the second term accounts for the impact of adversarial perturbations. Specifically, the second term scales as $2\epsilon/\sqrt{T}$ under $p=1$ attack, and as $2d\epsilon/\sqrt{T}$ under $p=\infty$ attack, where $d$ is the dimension of Hilbert space and $T$ is the number of training samples. Accordingly, our results bound the increase in sample complexity caused by the presence of an attacker, and they account for the power of the adversary via parameters $p$ and $\epsilon$. \\
\noindent  $\bullet$ {We study  a setting in which the classifier  is adversarially trained against a $p$-adversarial attack with $\epsilon$-perturbation budget, but it is  tested against a  $p'$-attack with $\epsilon'$-perturbation budget. We show that in the presence of this training-test mismatch, training with a strong adversary is the preferred strategy, as weak training adversaries may incur a positive non-vanishing term that scales as $d^{(1-1/p')}\epsilon'+d^{(1-1/p)}\epsilon.$}\\
\noindent  $\bullet$ Finally, we validate our main theoretical findings with numerical experiments.
\section{Problem Formulation}\label{sec:problemformulation}
In this section, we first introduce the quantum classification problem in the absence of quantum adversary, and define the conventional generalization error of a quantum classifier.  We then formulate the adversarial setting, and define the generalization error of an adversarially-trained classifier.

\subsection{Generalization Error of Quantum Classifiers}
As illustrated in Fig.~\ref{fig:1}(a),  a classical input $x$ is embedded into a quantum state $\rho(x)$ by a fixed and known \emph{quantum embedding} map $x \mapsto \rho(x)$.  The state $\rho(x)$ is a density matrix, i.e., a positive semi-definite, unit-trace matrix, defined in a finite-dimensional Hilbert space $\mathcal{H}$. Let $c \in \{1,\hdots,K\}$ denote the correct label assigned to input $x$ that takes values in one of the $K$ classes. The classical tuple $(x,c)$ is generated from an unknown data distribution $P(x,c)$. We assume $x$ to be discrete-valued to avoid some technicalities, but the analysis can be extended to continuous-valued inputs $x$.

The \textit{quantum  classifier} consists of a POVM applied to the quantum embedding $\rho(x)$. The POVM $\Pi =\{\Pi_c\}_{c=1}^K$ is defined by  positive semi-definite matrices $\Pi_c$, for $c=1,\hdots,K$, that satisfy the equality $\sum_{c=1}^K \Pi_c =I$, where $I$ denotes the identity matrix. We use $\mathcal{M}=\{\Pi: \Pi_c \geq 0,  \sum_{c=1}^K \Pi_c=I\}$ to denote the set of all POVMs.  By Born's rule,  a POVM $\Pi$ applied to a quantum state $\rho(x)$ yields the output class $c$ with probability ${\rm Tr}(\Pi_c \rho(x))$.

Accordingly, we consider as loss function the probability of error
\begin{align}
    \ell(\Pi,\rho(x), c) = 1-{\rm Tr} (\Pi_c \rho(x)) \label{eq:loss},
\end{align}
which is the probability of misclassifying state $\rho(x)$ given its true label $c$.  The goal of the quantum classification problem is to find the POVM $\Pi \in \mathcal{M}$ that minimizes the  \textit{population risk},
\begin{align}
    L(\Pi) = \Ebb_{P(x,c)}[\ell(\Pi,\rho(x), c)],
\end{align} which is the expected loss with respect to the distribution $P(x,c)$. 

However, the population risk cannot be evaluated by the classifier, since the data distribution $P(x,c)$ is unknown. Instead, the optimization of POVM is done with respect to the  \textit{empirical training risk},
\begin{align}
    \widehat{L}(\Pi,\mathcal{T}) = \frac{1}{T} \sum_{n=1}^T \ell(\Pi,\rho(x_n), c_n),
\end{align} which is evaluated using a training set $\mathcal{T}=\{(x_n,c_n)\}_{n=1}^T$ consisting of $T$ tuples $(x_n,c_n)$ generated i.i.d. from distribution $P(x,c)$. The difference between the population risk and the training risk is defined as the \textit{generalization error}
\begin{align}
  \mathcal{G}(\Pi,\mathcal{T}) = L(\Pi) -\widehat{L}(\Pi,\mathcal{T}) \label{eq:gen_error}
\end{align} obtained by the POVM $\Pi$.

\subsection{Adversarial Attacks on Quantum Classifiers}
In an adversarial setting, as illustrated in Fig.~\ref{fig:1}(b), a \textit{quantum adversary} can perturb the input quantum state $\rho(x)$ via the application of a completely positive trace preserving (CPTP) map, i.e., a quantum channel, with the aim of maximizing the classifier's loss \eqref{eq:loss}\cite{Lu_2020}. Targeting a worst-case scenario, the adversary is assumed to know the quantum classifier $\Pi$,  the loss function \eqref{eq:loss}, as well as the quantum embedding map $x \mapsto \rho(x)$, resulting in  \textit{white-box} attacks.

To define the power of the adversary, we constrain the distance between the density matrices before and after the perturbation. To this end, we adopt the $p$-Schatten norm. For two density matrices $\rho_1$ and  $\rho_2$ and $p \in [1, \infty)$, the \emph{$p$-Schatten  distance} $D_p(\rho_1, \rho_2)$ is defined as \begin{equation}D_p(\rho_1, \rho_2)=\Vert \rho_1 -\rho_2 \Vert_p = ({\rm Tr}(|\bar{\rho}|^p))^{1/p},\end{equation} where $\bar{\rho}=\rho_1-\rho_2$ and  $|\bar{\rho}|=\sqrt{\bar{\rho} \bar{\rho}^{\dag}}$. In the limiting case of $p=\infty$, the distance $D_\infty(\rho_1, \rho_2)$ is defined as $D_\infty(\rho_1, \rho_2) = \max(\{\alpha(|\bar{\rho}|)\})$ where $\{\alpha(|\bar{\rho}|)\}$ is the set of eigenvalues of $|\bar{\rho}|$.

A \emph{$p$-adversarial attack with a perturbation budget  $\epsilon\geq0$}  can produce any quantum state $\lambda$ satisfying $D_p(\rho(x),\lambda )\leq \epsilon$. Assuming that the adversary maximizes the loss $\ell(\Pi,\lambda,c)$ incurred by the quantum classifier under this perturbation budget, the resulting \textit{adversarial loss} of the classifier $\Pi$ on  data tuple $(\rho(x),c)$ is given as
\begin{align}
    \ell_{p,\epsilon} (\Pi,\rho(x), c) & = \underset{\lambda: D_p(\rho(x),\lambda ) \leq \epsilon}{\max} \ell(\Pi,\lambda,c)\label{eq:adversarial_loss},
\end{align}  where $\ell(\Pi,\lambda,c)$ is defined as in \eqref{eq:loss}.  
In this paper, we will focus on the extreme cases with $p=1$ and $p=\infty$ adversarial attacks.

In   the presence of a $p$-adversarial attack with perturbation budget $\epsilon$, the performance of the quantum classifier is measured by   the \textit{adversarial population risk}
\begin{align}
    L_{p,\epsilon}(\Pi) = \Ebb_{P(x,c)}[\ell_{p,\epsilon} (\Pi,\rho(x), c)], \label{eq:adversarial_population}
\end{align} which is the expected adversarial loss with respect to the unknown distribution $P(x,c)$. 

\subsection{Generalization Error of Adversarially Trained Classifiers}

Suppose that the quantum classifier is aware of the presence of a $p$-adversarial attack with perturbation budget $\epsilon$. While the adversarial population risk cannot be directly evaluated, the quantum classifier can be trained by optimizing the \textit{adversarial training risk}
\begin{align}
     \widehat{L}_{p,\epsilon}(\Pi,\mathcal{T}) = \frac{1}{T}\sum_{n=1}^T \ell_{p,\epsilon} (\Pi, \rho(x_n), c_n),\label{eq:adversarial_training}
\end{align} which is the empirical average of the adversarial loss \eqref{eq:adversarial_loss} over the training set $\mathcal{T}$. This results in a min-max optimization problem with the outer minimization over POVMs and the inner maximization over perturbations of quantum states \cite{Lu_2020}. 

In this work, we are interested in characterizing the \textit{adversarial generalization error}. The adversarial generalization error of a POVM $\Pi$ is the difference between adversarial population risk \eqref{eq:adversarial_population} and adversarial training loss \eqref{eq:adversarial_training}, i.e.,
\begin{align}
    \mathcal{G}_{p,\epsilon}(\Pi,\mathcal{T}) = L_{p,\epsilon}(\Pi)-\widehat{L}_{p,\epsilon}(\Pi,\mathcal{T}). \label{eq:adversarial_generror}
\end{align} Note that in the limit as $\epsilon \rightarrow 0$, the adversarial generalization error $\mathcal{G}_{p,\epsilon}(\Pi,\mathcal{T})$ coincides with the standard generalization error $\mathcal{G}(\Pi,\mathcal{T})$ in \eqref{eq:gen_error}.

\section{Preliminaries}\label{sec:background}

In this section, we first present the main result of \cite{banchi2021generalization}, which gives a high-probability, information-theoretic, upper bound on the generalization error \eqref{eq:gen_error} for conventional quantum learning. We then outline the key steps in the derivation of the upper bound, which will be useful in the next section to derive the proposed upper bounds on the adversarial generalization error.

\begin{theorem}[Banchi \emph{et. al} \cite{banchi2021generalization}] \label{gener}
For any POVM $\Pi \in \mathcal{M}$, the following upper bound on the generalization error $\mathcal{G}(\Pi,\mathcal{T})$ holds with
probability at least $1-\delta$, for $\delta \in (0,1)$, with respect to random draws of of the training set $\mathcal{T}$,
    \begin{align}
        \mathcal{G}(\Pi,\mathcal{T}) & \leq 2\sqrt{\frac{ 2^{I_{2}(X:Q)}K}{T}} + \sqrt{\frac{2 \log(2/\delta)}{T}} := \mathcal{B}, \label{eq:banchi_result}
    \end{align}
where $I_2(X:Q)$ denotes the 2-Renyi mutual information between the quantum state space $Q$ and the classical feature space $X$ under state $\rho^{XQ}=\sum_x P(x) |x\rangle\langle x|\otimes \rho(x)$, which is given by \begin{equation}{I_{2}(X:Q)}=2 \log_2\Bigg({\rm Tr} \sqrt{\sum_x P(x)\rho(x)^2}\Bigg). \label{eq:renyimi}\end{equation}
\end{theorem} 
The derivation of the upper bound in \eqref{eq:banchi_result} follows two main steps. 
In the first step, the generalization error $\mathcal{G}(\Pi,\mathcal{T})$ of a POVM $\Pi \in \mathcal{M}$  is upper bounded as
\begin{align}
  \mathcal{G}(\Pi,\mathcal{T}) \leq \sup_{\Pi \in \mathcal{M}} | L(\Pi) -\widehat{L}(\Pi,\mathcal{T})|: = \mathcal{U}(\mathcal{M},\mathcal{T}) \label{eq:intermediate_ub},
\end{align}
where $\mathcal{U}(\mathcal{M},\mathcal{T})$ denotes the \textit{uniform deviation bound} that depends on the training set $\mathcal{T}$ and the set $\mathcal{M}$ of POVMs. In the second step, the uniform deviation bound is upper bounded by leveraging a classical result from statistical learning theory. This result, stated next, hinges on the fact that the loss function  in \eqref{eq:loss} satisfies the inequality $0 \leq \ell(\cdot,\cdot,\cdot) \leq 1$. 
\begin{lemma}[Shalev-Schwartz and Ben David\cite{shalev2014understanding}]\label{lem:rademacher}
    With probability at least $1-\delta$, for $\delta \in (0,1)$, with respect to random draws of the training set $\mathcal{T}$, the following inequality holds
\begin{align}
    \mathcal{U}(\mathcal{M},\mathcal{T}) \leq 2\mathcal{R}(\mathcal{M}) + \sqrt{\frac{2 \log (2/\delta)}{T}}, \label{eq:nonadversarial_udb}
\end{align} where 
\begin{align}
    \mathcal{R}(\mathcal{M}) = \Ebb_{\mathcal{T}} \Ebb_{\boldsymbol{\sigma}}\biggl[\sup_{\Pi \in \mathcal{M}} \frac{1}{T} \sum_{n=1}^{T}\hspace{-0.1cm}\sigma_n \ell(\Pi,\rho(x_n),c_n) \biggr] \label{eq:rademacher}
\end{align} is the \textit{Rademacher complexity} of the set $\mathcal{M}$ of POVMs. In  \eqref{eq:rademacher},  $\boldsymbol{\sigma}=(\sigma_1,\hdots,\sigma_T)$ denotes a vector of $T$ i.i.d Rademacher variables $\sigma_i$ that takes value $\pm 1$ with equal probability.
\end{lemma}
An information-theoretic characterization of the Rademacher complexity \eqref{eq:rademacher}
then yields the upper bound in \eqref{eq:banchi_result}.

\section{Generalization Bounds for Adversarially Trained Quantum Classifiers}
In this section, we present our main results, which provides  information-theoretic upper bounds on the adversarial generalization error $\mathcal{G}_{p,\epsilon}(\Pi,\mathcal{T})$ defined in \eqref{eq:adversarial_generror}.

\subsection{Key Technical Challenge}
To derive upper bounds on the adversarial generalization error $\mathcal{G}_{p,\epsilon}(\Pi,\mathcal{T})$, one can follow similar steps as discussed in Sec.~\ref{sec:background}, targeting the  \textit{adversarial uniform deviation bound}   \begin{equation}\mathcal{U}_{p,\epsilon}(\mathcal{M},\mathcal{T})= \sup_{\Pi \in \mathcal{M}} | L_{p,\epsilon}(\Pi) -\widehat{L}_{p,\epsilon}(\Pi,\mathcal{T})|\end{equation} on the adversarial generalization error $\mathcal{G}_{p,\epsilon}(\Pi,\mathcal{T})$. The uniform deviation bound can be further upper bounded, as in Lemma~\ref{lem:rademacher}, as a function of the  \textit{adversarial Rademacher complexity} \begin{align}
    \mathcal{R}_{p,\epsilon}(\mathcal{M})\hspace{-0.1cm} =\hspace{-0.1cm} \Ebb_{\mathcal{T}} \Ebb_{\boldsymbol{\sigma}}\biggl[\sup_{\Pi \in \mathcal{M}} \frac{1}{T} \sum_{n=1}^{T} \hspace{-0.1cm}\sigma_n \ell_{p, \epsilon}(\Pi,\rho(x_n),c_n) \biggr] \label{eq:adversarial_rademacher}. 
\end{align} Specifically, as in  Lemma~\ref{lem:rademacher}, with probability at least $1-\delta$, for $\delta \in (0,1)$, the following inequality holds \begin{align}\label{eq:adversarial_udb}
    \mathcal{U}_{p,\epsilon}(\mathcal{M},\mathcal{T}) \leq 2\mathcal{R}_{p,\epsilon}(\mathcal{M}) + \sqrt{\frac{2 \log(2/\delta)}{T}}. 
\end{align}

However, evaluating the adversarial Rademacher complexity $\mathcal{R}_{p,\epsilon}(\mathcal{M})$ is challenging. The function $\mathcal{R}_{p,\epsilon}(\mathcal{M})$ in \eqref{eq:adversarial_rademacher} is  defined using the adversarial loss $\ell_{p, \epsilon}(\Pi,\rho(x),c)$, which entails  a  maximization problem over the set $\{\lambda: \lambda \succeq 0, {\rm Tr}(\lambda)=1, \Vert \rho(x)- \lambda \Vert_p \leq \epsilon\}$ of density matrices  that satisfy the perturbation constraint. In  the corresponding problem   studied in  \cite{yin2019rademacher} for classical adversarial learning,   the relevant constraint imposes a bound on the $l_{\infty}$-norm based perturbation of the classical input, and the resulting adversarial loss can be easily evaluated in closed form. In contrast, the constrained optimization underlying the quantum adversarial loss $\ell_{p, \epsilon}(\Pi,\rho(x),c)$ appears not to admit a closed-form solution in general.


\subsection{Main Results}
To state the main results, we make the following assumption.
\begin{restatable}{assumption}{asegval}\label{asu1}
    The quantum embedding map $x \mapsto \rho(x)$ from classical input $x$ to density matrix $\rho(x)$ is such that the minimum eigenvalue of the density matrix $\rho(x)$ satisfies the inequality $\alpha_{\min}(\rho(x))\geq \Delta$ for some $\Delta \in (0,1/d]$, where $d$ is the dimension of the Hilbert space.
\end{restatable}

Assumption 1 imposes a constraint on the entropy   of the quantum embedding, requiring all quantum states $\rho(x)$ to have all non-zero eigenvalues, and hence maximum R\'enyi entropy of order zero \cite{muller2013quantum}. In practice, the quantum embedding may be noisy, which is modelled by a CPTP map $\mathcal{E}(\cdot)$, whereby the input classical data $x$ is mapped to a noisy state $\rho'(x)$ as $x \mapsto \mathcal{E}(\rho(x))=\rho'(x)$.
The minimal eigenvalue of the resulting noisy state $\rho'(x)$ is greater than or equal to that of the clean state, i.e., 
$  \alpha_{\min}(\mathcal{E}(\rho(x))) \geq \alpha_{\min}(\rho(x)).$
Thus,
 noisy quantum states can satisfy Assumption~\ref{asu1} even when corresponding clean states don't.

The following theorem gives an upper bound on the adversarial generalization error $\mathcal{G}_{1,\epsilon}(\Pi)$ defined in \eqref{eq:adversarial_generror} with $p=1$. 
\begin{theorem}\label{thm:1}
  Assume that the $K$ classes are equi-probable and that we have a $p=1$-adversarial attack with a perturbation budget   $\epsilon \leq 2\Delta$.  Under Assumption~\ref{asu1}, the following upper bound on the adversarial generalization error $\mathcal{G}_{1,\epsilon}(\Pi,\mathcal{T})$,  for any POVM $\Pi \in \mathcal{M}$, holds with probability at least $1-\delta$, for $\delta \in (0,1)$, 
    \begin{gather}
    \mathcal{G}_{1,\epsilon}(\Pi, \mathcal{T})\leq \mathcal{B}+ 2\sqrt{\frac{K}{T}}\epsilon, \label{eq:main_result1}
    \end{gather} where $\mathcal{B}$ is as defined in \eqref{eq:banchi_result}.
    \end{theorem}

The bound in \eqref{eq:main_result1} shows that  the adversarial generalization error can be upper bounded in terms of the non-adversarial generalization bound \eqref{eq:banchi_result} with an additional  term that is directly proportional to the perturbation budget $\epsilon$. This  term quantifies the impact of the adversarial perturbation on generalization,  and it recovers the bound  \eqref{eq:banchi_result} for  $\epsilon=0$. Furthermore, by the upper bound in \eqref{eq:main_result1},  in the limit of a large number of observations $T$, the adversarial generalization error vanishes. These results hold under the constraint $\epsilon \leq 2 \Delta$ on the power of the adversary, which is more restrictive for less noisy quantum embeddings $\rho(x)$ with a smaller minimum eigenvalue $\alpha_{\min}(\rho(x))$. 


We now present an upper bound on the adversarial 
generalization error $\mathcal{G}_{\infty, \epsilon}(\Pi,\mathcal{T})$ under $\infty$-Schatten norm attacks.
\begin{theorem}\label{thm:2}
      Assume that the $K$ classes are equi-probable and that we have a $p=\infty$-adversarial attack with a perturbation budget   $\epsilon \leq \Delta$.  Under Assumption~\ref{asu1}, the following upper bound on the adversarial generalization error $\mathcal{G}_{\infty,\epsilon}(\Pi,\mathcal{T})$,  for any POVM $\Pi \in \mathcal{M}$, holds with probability at least $1-\delta$, for $\delta \in (0,1)$,
    \begin{gather}
    \mathcal{G}_{\infty,\epsilon}(\Pi, \mathcal{T})\leq \mathcal{B}+ 2d\sqrt{\frac{K}{T}}\epsilon, \label{eq:main_result2}
    \end{gather} where $\mathcal{B}$ is as defined in \eqref{eq:banchi_result}.
\end{theorem}
 For any given perturbation level $\epsilon$, $p$-adversarial attacks with $p=\infty$ are stronger than with $p=1$, since they allow for  perturbations in a larger volume of the Hilbert space. In a manner consistent with this observation, the additional term in the bound \eqref{eq:main_result2} is larger than in \eqref{eq:main_result1}, with the relative increase factor  equal to the dimension $d$  of the Hilbert space. The result holds under the more restrictive assumption $\epsilon \leq \Delta $.


The generalization bounds derived in the previous two theorems vanish in the limit of a large number of samples,  $T \rightarrow \infty$. These results hold under  Assumption~\ref{asu1}, which requires the quantum embeddings to be sufficiently noisy, and on the stated upper bounds for the perturbation $\epsilon$. As we show next, even when removing these assumptions, it is possible to show that the adversarial generalization error is given by the adversarial generalization bound \eqref{eq:banchi_result} with the addition of a term proportional to the perturbation level $\epsilon$. However, these additional terms do not vanish as $T$ increases. We leave it as an open problem to establish tighter bounds in this regime.  
\begin{theorem}\label{thm:generalresult}
  Assume that the $K$ classes are equi-probable and that we have a $p$-adversarial attack with any perturbation budget   $\epsilon\geq 0$. For any POVM $\Pi \in \mathcal{M}$,  the following upper bound on the adversarial generalization error $\mathcal{G}_{p,\epsilon}(\Pi,\mathcal{T})$ holds with probability at least $1-\delta$, for $\delta \in (0,1)$,
\begin{align}
    \mathcal{G}_{p, \epsilon}(\Pi, \mathcal{T}) \leq \mathcal{B} + \begin{cases}  \epsilon\sqrt{2 d(1 + \frac{K-1}{T})}, & \text{$p=1$} \\ 2\epsilon d\sqrt{(1 + \frac{K-1}{T})}, & \text{$p=\infty$}.
    \end{cases} \label{eq:general_result}
\end{align}
\end{theorem}
\section{Generalization Bounds Under Adversarial Mismatch}
In the previous sections, we have considered the setting in which the quantum classifier is trained by assuming the same type of attacks encountered during testing.  This is seldom true in practice: a quantum classifier $\Pi$ adversarially trained against $p$-adversarial attacks with an $\epsilon$-perturbation budget can encounter a generally different $p'$-adversarial attack with $\epsilon'$-budget during testing. In this section, we quantify the adversarial generalization error under adversarial mismatch.

We define the \textit{mismatched adversarial generalization error},  \begin{align*}
\mathcal{G}_{p,p',\epsilon,\epsilon'}(\Pi,\mathcal{T}) = L_{p',\epsilon'}(\Pi)-\widehat{L}_{p,\epsilon}(\Pi,\mathcal{T}),
\end{align*} of a POVM $\Pi$ as the difference between the adversarial population risk $L_{p',\epsilon'}(\Pi)$, evaluated under $p'$-adversarial attack with $\epsilon'$-perturbation budget, and the adversarial training risk $L_{p,\epsilon}(\Pi)$, evaluated under $p$-adversarial attack with $\epsilon$-perturbation budget. 
To characterize the mismatched adversarial generalization error as a function of the  generalization error $\mathcal{G}_{p,\epsilon}(\Pi,\mathcal{T})$, we first define the following notion of relative strength of the adversaries.
\begin{definition} \label{def:strenght}
  A $p$-adversarial attack with perturbation budget $\epsilon \geq 0$ is said to be stronger than a $p'$-adversarial attack with perturbation budget $\epsilon' \geq 0$  if the following inclusion condition \begin{equation}\label{eq:stronger}\{\lambda:D_p(\lambda, \rho)\leq\epsilon\} \supset \{\lambda:D_{p'}(\lambda, \rho)\leq\epsilon'\} \end{equation} holds for all density matrices $\rho$. In this case, we also say that the second attack is weaker than the first.

\end{definition}
The definition above is justified by the fact that a stronger attack, satisfying condition (\ref{eq:stronger}), would be able to further increase the adversarial loss (\ref{eq:adversarial_loss}) as compared to a weaker attack.  
The following lemma provides sufficient  conditions that  guarantee an adversary to be stronger than another.
\begin{lemma}\label{lem:strength}
     A $p$-adversarial attack with budget $\epsilon>0$ is stronger than a $p'$-adversarial attack with budget $\epsilon'>0$  if 
    \begin{align}
        \epsilon'< \begin{cases}
        d^{1/p'-1/p}\epsilon, & p\leq p'\nonumber\\ 2 d^{1/p'-1/p-1}\epsilon, & p> p'
        \end{cases}
    \end{align}
\end{lemma}
With these definitions, we have the following result. 
\begin{theorem} \label{thm:mismatched}
 Assume that the quantum classifier is adversarially trained assuming a $p$-adversarial attack with  perturbation budget   $\epsilon\geq 0$, while a $p'$-adversarial attack with perturbation budget   $\epsilon'\geq 0$ affects the quantum embeddings during testing. If the training adversarial attack is stronger than the testing adversarial attack, the following relation holds,
\begin{align}\label{eq:mis_leq}
\mathcal{G}_{p,\epsilon}(\Pi,\mathcal{T}) - \xi \leq\mathcal{G}_{p,p'\epsilon,\epsilon'} (\Pi,\mathcal{T})\leq \mathcal{G}_{p,\epsilon}(\Pi,\mathcal{T}),
\end{align} 
where  $$\xi=d^{(1-1/p')}\epsilon'+d^{(1-1/p)}\epsilon$$ is a function of the parameters $(p,p',\epsilon,\epsilon')$. If the training adversary is weaker than the testing adversary, we have 
\begin{align}\label{eq:mis_geq}
\mathcal{G}_{p,\epsilon}(\Pi,\mathcal{T}) \leq \mathcal{G}_{p,p'\epsilon,\epsilon'} (\Pi,\mathcal{T}) \leq \mathcal{G}_{p,\epsilon}(\Pi,\mathcal{T}) + \xi.
\end{align}
\end{theorem}
Theorem~\ref{thm:mismatched} gives insights on how best to adversarially train the classifier so that it generalizes well when tested against a possibly different adversary. {In particular, the upper bound  \eqref{eq:mis_leq} guarantees that if the training adversary is \textit{stronger} than the testing adversary, the mismatched generalization error is no larger  than the generalization error $\mathcal{G}_{p,\epsilon}(\Pi,\mathcal{T})$ obtained when  the stronger attacker is also present at test time. From Lemma \ref{lem:strength}, a way to ensure a stronger attacker at training time is to train assuming $p=\infty$ and a sufficiently large $\epsilon$. Conversely, by \eqref{eq:mis_geq}, assuming a weaker adversary during training yields a mismatched generalization error that can exceed  the generalization error $\mathcal{G}_{p,\epsilon}(\Pi,\mathcal{T})$ with the weaker test-time attacker by a  non-vanishing (with $T$) term  $\xi.$

\section {Examples and Final Remarks}
\begin{figure}[t!]
    \begin{minipage}{0.48\columnwidth}
\includegraphics[scale=0.29, clip=true, trim =0.4in 0in 0.4in 0.5in]{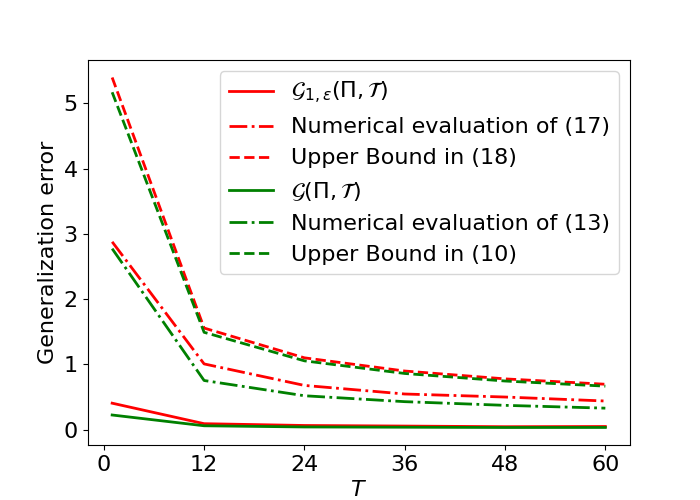} 
  \end{minipage}
    \begin{minipage}{0.47\columnwidth}
\includegraphics[scale=0.29,clip=true, trim =0.6in 0in 0.65in 0.5in]{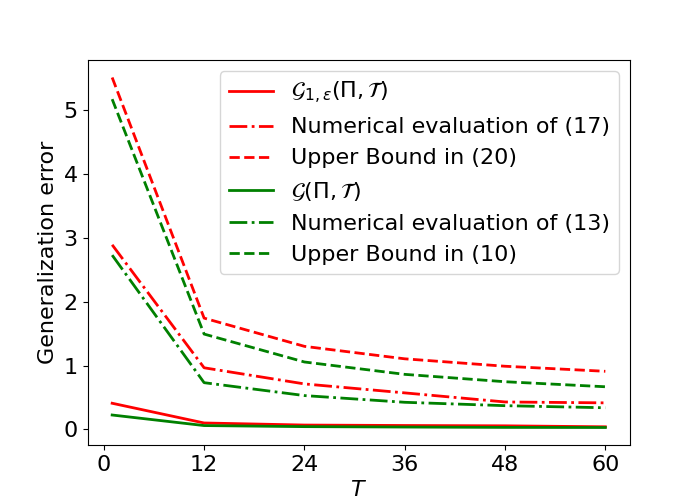} 
    \end{minipage}
  \caption{ True generalization errors for non-adversarial ($\mathcal{G}(\Pi,\mathcal{T})$) and adversarial ($\mathcal{G}_{1,\epsilon}(\Pi,\mathcal{T})$) settings, compared with numerically evaluated uniform deviation bounds \eqref{eq:nonadversarial_udb} and \eqref{eq:adversarial_udb} and derived bounds as a function of the training set size $T$:  (left) $\epsilon=0.08 \leq 2 \Delta=0.1$, and (right) $\epsilon=0.12 >2\Delta$.}
  \label{fig:expt1}
\end{figure}
We consider a quantum binary classification problem with equi-probable class labels $c \in \{0, 1\}$. For each class $c$, we obtain the discrete-valued input $x$ by finely quantizing a continuous-valued feature input $ \tilde{x} \in \mathbb{R}$ so that the discrete sum in \eqref{eq:renyimi} can be evaluated via numerical integration \cite{banchi2021generalization}. The input $\tilde{x}$ is sampled from the conditional Gaussian distribution $P(x|c)=\mathcal{N}(\mu_c, 1)$ with mean $ \mu_c = (-1)^c$. We consider a depolarized quantum embedding, with noise strength $q \in (0,1)$, that maps $x$ to the quantum state $\rho(x) = (1-q)\vert x \rangle \langle x \vert + qI/d$, where the pure state $\vert x \rangle$ is obtained as
\begin{gather}
    \vert x \rangle = U_\theta(x) \vert 0 \rangle, \; U_\theta(x) = R_X(x){\rm Rot}_\theta R_X(x),
\end{gather}
with $\theta = (\theta_1,\theta_2,\theta_3) \in [0, 2\pi)^3.$ Here, $ {\rm Rot}_{\theta} = \exp(-i\Vec{\theta} \cdot \Vec{\sigma})$ and $R_X(x) = {\rm Rot}_{(x,0,0)}$ are single qubit rotation gates,  where $\Vec{\sigma} = (\sigma_x, \sigma_y, \sigma_z)$ denotes the vector of the Pauli matrices. In our experiments, we fix $\theta = (\pi/4,\pi/4,\pi/4)$ and $q = 0.05$, which results in $\Delta=0.05$.

In Fig.~\ref{fig:expt1}, we plot the true non-adversarial and adversarial generalization errors, i.e., $\mathcal{G}(\Pi, \mathcal{T})$ and $\mathcal{G}_{1,\epsilon}(\Pi, \mathcal{T})$ (for $p=1$)  respectively, for the POVM $\Pi = \{\vert 0 \rangle\langle 0\vert,\vert 1 \rangle\langle 1\vert\}$  when $\epsilon \leq 2\Delta$ (left) and $\epsilon >2\Delta$ (right) as a function of the training set size $T$. To validate our analysis, we also evaluate numerically the Rademacher complexity based uniform deviation bounds \eqref{eq:nonadversarial_udb} and \eqref{eq:adversarial_udb} for non-adversarial and adversarial errors with $\delta=0.8$; and  we plot the derived adversarial upper bounds \eqref{eq:main_result1} (left) and \eqref{eq:general_result} (right), along with the non-adversarial bound in \eqref{eq:banchi_result}.

The true generalization bounds follow a similar trend in both plots, with the adversarial generalization error being larger than the non-adversarial counterpart, and with both errors tending to $0$  for large values of the data set size $T$. Furthermore, when the adversary's perturbation is limited as  $\epsilon \leq 2\Delta$, this behaviour is reproduced by the derived upper bound in Theorem~\ref{thm:1}. From the uniform deviation bounds it can be seen that the adversarial Rademacher complexity exceeds the non-adversarial Rademacher complexity. For the case when $\epsilon>2\Delta$, i.e., when Assumption~\ref{asu1} is not satisfied,  while capturing the general decrease with $T$ of the generalization error, our bound \eqref{eq:general_result} is loose. We leave it as an open problem to derive tighter bounds in this regime. This observation also suggests that Assumption~\ref{asu1} is only instrumental in facilitating the 
derivation of the bound, which requires the optimization over the attacker's channel,  rather than indicating a ``phase transition" in the generalization behavior. 

\section*{Acknowledgments}
The work of OS was supported by an Open Fellowship of the EPSRC with reference EP/W024101/1, by the EPSRC project EP/X011852/1, and by the European Union’s Horizon Europe Project CENTRIC under Grant 101096379.
PG wishes to thank Mr. Charalampos Perdikis for the many useful discussions about code optimization and the use of multiprocessing and multithreading in accelerating his code.

\clearpage

\bibliographystyle{IEEEtran}
\bibliography{refs}

\clearpage
\appendices
\section{Proof of Theorem~\ref{thm:1} and Theorem~\ref{thm:2}}\label{app:theorems_12}
The key idea of the proofs is to upper bound the adversarial Rademacher complexity  \eqref{eq:adversarial_rademacher} via  the non-adversarial Rademacher complexity \eqref{eq:rademacher} and an additional term that accounts for the impact of perturbation. To this end, we equivalently write the adversarial Rademacher complexity \eqref{eq:adversarial_rademacher} as
\begin{align}
    &\mathcal{R}_{p,\epsilon}(\mathcal{M}) = \Ebb_{\mathcal{T} } \Ebb_{\boldsymbol{\sigma}}\biggl[ \sup_{\Pi \in \mathcal{M}} \frac{1}{T} \sum_{n=1}^{T}\sigma_n \times \nonumber \\&   \max_{\lambda:D_p(\lambda,\rho(x_n)) \leq \epsilon} \Bigl(1 - {\rm Tr}( \Pi_{c_n}\rho(x_n)) - {\rm Tr}( \Pi_{c_n}(\lambda - \rho(x_n))) \Bigr) \biggr]. \label{eq:proof_1}
\end{align}Using the inequality   $\sup(f+g) \leq \sup f +\sup g$ in the upper bound \eqref{eq:proof_1}, we obtain
\begin{align}
    &\mathcal{R}_{p,\epsilon}(\mathcal{M}) \leq \mathcal{R}(\mathcal{M}) +\Delta\mathcal{R}_{p,\epsilon}(\mathcal{M}), \label{eq:decomposition}
    \end{align}where $\mathcal{R}(\mathcal{M})$ is as defined in \eqref{eq:rademacher}, and 
   \begin{align*}&\Delta\mathcal{R}_{p,\epsilon}(\mathcal{M})= \Ebb_{\mathcal{T}} \Ebb_{\boldsymbol{\sigma}}\biggl[ \sup_{\Pi \in \mathcal{M}} \frac{1}{T} \sum_{n=1}^{T}\sigma_n \times \\&  \max_{\lambda:D_p(\lambda,\rho(x_n)) \leq \epsilon} {\rm Tr}\Bigl( \Pi_{c_n}( \rho(x_n)-\lambda) \Bigr) \biggr]
\end{align*} may be defined as the perturbation Rademacher complexity.

We continue by writing the POVM elements $\Pi_c$ in terms of their eigendecomposition as $\Pi_c = U_c \bar{\Pi}_c U_c^\dagger$, where $\bar{\Pi}_c$ denotes the diagonal matrix of eigenvalues. Using the cyclic property of the trace, $\Delta\mathcal{R}_{p,\epsilon}(\mathcal{M})$ can be equivalently written as
\begin{align}    
    \Delta\mathcal{R}_{p,\epsilon}(\mathcal{M}) &=\Ebb_{\mathcal{T} } \Ebb_{\boldsymbol{\sigma}}\biggl[ \sup_{\Pi \in \mathcal{M}} \frac{1}{T} \sum_{n=1}^{T}\sigma_n \times \nonumber \\& \max_{\lambda:D_p(\lambda,\rho(x_n)) \leq \epsilon} {\rm Tr}\Bigl( \bar{\Pi}_{c_n}U^\dagger_{c_n} (\rho(x_n) - \lambda) U_{c_n} \Bigr) \biggr] \nonumber \\
    &= \Ebb_{\mathcal{T} } \Ebb_{\boldsymbol{\sigma}}\biggl[ \sup_{\Pi \in \mathcal{M}} \frac{1}{T} \sum_{c=1}^K \sum_{n=1}^{T} \mathds{1}(c_n=c)\sigma_n \times \nonumber \\& \max_{\lambda:D_p(\lambda,\rho(x_n)) \leq \epsilon} {\rm Tr}( \bar{\Pi}_{c}\bar{\tau}(x_n,c) \Bigr) \biggr] \label{eq:proof_12},  
\end{align}
where we have defined $\bar{\tau}(x_n,c) = U_c^\dagger (\rho(x_n) - \lambda) U_c$.

We now proceed to upper bound $\Delta\mathcal{R}_{p,\epsilon}(\mathcal{M})$ for the case of $p=1$, which gives the required upper bound in Theorem~\ref{thm:1}.

\subsection{$D_{1}(\cdot, \cdot)$ perturbation Rademacher complexity}
For fixed $c$, the inner maximization in \eqref{eq:proof_12} is achieved when $\bar{\tau}(x_n,c)$ is diagonal with entries 
\begin{align} \label{eq:pert_matr}
    (\bar{\tau}(x_n,c))_{ii} = \begin{cases}
        +\frac{\epsilon}{2} & \text{if $(\bar{\Pi}_{c})_{ii} = \alpha_{\max}(\Pi_{c})$}\\
        -\frac{\epsilon}{2} & \text{if $(\bar{\Pi}_{c})_{ii} = \alpha_{\min}(\Pi_{c})$}\\
        0 & \text{otherwise},
    \end{cases}
\end{align} where $\alpha_{\max}(\cdot)$ and $\alpha_{\min}(\cdot)$ respectively denote the maximum and minimum eigenvalues of `$\cdot$'.
It can be verified that this choice of $\bar{\tau}(x_n,c)$ yields a physical density matrix $\lambda$. In particular, the condition $\epsilon \leq 2\Delta$ guarantees that the minimal eigenvalue of $\lambda$ is positive (for 2 linear operators $A$, $B$, we have $\alpha_{\min}(A-B) \geq \alpha_{\min}(A) - \alpha_{\max}(B)$).

Now, defining $
    Q_{\boldsymbol{\sigma}, c} = \sum_{n=1}^T \sigma_n \mathds{1}(c_n = c) \bar{\tau}(x_n,c),
$
 we can re-write \eqref{eq:proof_12} as
\begin{align*}
    \Delta\mathcal{R}_{1,\epsilon}(\mathcal{M}) = \Ebb_{\mathcal{T}}{\mathbb{E}}_{\boldsymbol{\sigma}}\Bigl[\sup_{\Pi} \frac{1}{T} \sum_c {\rm Tr}( \bar{\Pi}_c   Q_{\boldsymbol{\sigma}, c}) \Bigr].
\end{align*} Applying H{\"o}lder's inequality, we get the relation ${\rm Tr}( \bar{\Pi}_c   Q_{\boldsymbol{\sigma}, c})\leq \Vert   Q_{\boldsymbol{\sigma}, c} \Vert_1 \Vert \bar{\Pi}_c \Vert_{\infty} \leq \Vert   Q_{\boldsymbol{\sigma}, c} \Vert_1  $ since 
$\lVert \bar{\Pi_c} \rVert_\infty \leq 1$. Subsequently, we have
\begin{align*}
    \Delta\mathcal{R}_{1,\epsilon}(\mathcal{M}) \leq \Ebb_{\mathcal{T} }{\mathbb{E}}_{\boldsymbol{\sigma}}\Bigl[\sup_{\Pi} \frac{1}{T} \sum_c \Vert Q_{\boldsymbol{\sigma}, c}\Vert_1 \Bigr]. 
 \end{align*} Since 
 $Q_{\boldsymbol{\sigma}, c}$ is  diagonal, the trace norm $\Vert Q_{\boldsymbol{\sigma}, c} \Vert_1$ evaluates as the sum of the absolute values of its diagonal elements. We thus have
\begin{align}
    \Delta\mathcal{R}_{1,\epsilon}(\mathcal{M}) \leq \frac{2}{T}{\mathbb{E}}_{\mathcal{T}} \Ebb_{\boldsymbol{\sigma}}\biggl[ \sum_c  \Bigl | \sum_{n=1}^T \frac{\epsilon}{2}\sigma_n \mathds{1}(c_n = c) \Bigr | \biggr]. \label{eq:proof_13} 
\end{align} Let $T_c$ denote the number of examples in the training set $\mathcal{T}$ that belongs to class $c$. Then,
for $K$ equiprobable classes, the upper bound \eqref{eq:proof_13} evaluates as
\begin{align*}
    \Delta\mathcal{R}_{1,\epsilon}(\mathcal{M}) \leq \frac{K}{T}{\mathbb{E}}_{\mathcal{T}} \Ebb_{\boldsymbol{\sigma}}\Ebb_c \biggl[   \Bigl| \sum_{n=1}^{T_c} \epsilon\sigma_n  \Bigr| \biggr].
\end{align*}
Using Khintchine's inequality (see, e.g., \cite{haagerup1981best}), we have $\Ebb_{\boldsymbol{\sigma}}[| \sum_{n=1}^{T_c} \epsilon\sigma_n |] \leq \epsilon \sqrt{T_c} $. This results in
\begin{gather*}
    \Delta\mathcal{R}_{1,\epsilon}(\mathcal{M}) \leq \epsilon\frac{K}{T}\Ebb_{\mathcal{T}}\Ebb_c  [\sqrt{T_c}] \leq\epsilon\frac{K}{T} \sqrt{\Ebb_{\mathcal{T}}\Ebb_c [T_c]},
\end{gather*}
where the last inequality is due to Jensen's inequality. 
Finally, noting that the classes are equi-probable, the 
expected value of $T_c$ evaluates as $T/K$, yielding
\begin{gather*}
    \Delta\mathcal{R}_{1,\epsilon}(\mathcal{M}) \leq \epsilon\sqrt{\frac{K}{T}}.
\end{gather*}
Using this in \eqref{eq:decomposition}, together with the upper bound in \eqref{eq:banchi_result} returns the upper bound of \eqref{eq:main_result1}.

We now upper bound $\Delta \mathcal{R}_{p,\epsilon}(\mathcal{M})$ for $p=\infty$, which gives the required upper bound in Theorem~\ref{thm:2}.
\subsection{$D_{\infty}(\cdot, \cdot)$ perturbation Rademacher complexity}

To upper bound $\Delta \mathcal{R}_{p,\epsilon}(\mathcal{M})$  in \eqref{eq:proof_12}, we start by arranging the 
set $\{\alpha(\Pi_c)_i\}$ of eigenvalues of $\Pi_c$ in increasing order in $i$.  Then, define the median eigenvalue as
\begin{align*}
    \alpha_{{\rm med}} = \begin{cases}\frac{\alpha(\Pi_c)_{d/2} + \alpha(\Pi_c)_{d/2+1}}2 & \mbox{if} \hspace{0.1cm}  d \hspace{-0.2cm}\mod 2 = 0 \\
    \alpha(\Pi_c)_{\lceil d/2 \rceil} & \mbox{if} \hspace{0.1cm} d \hspace{-0.2cm} \mod 2 = 1.
    \end{cases}
\end{align*}
For fixed $c$, the inner maximization in (\ref{eq:proof_12}) is achieved when $\bar{\tau}(x_n, c)$ is diagonal with entries 
\begin{align*}
    (\bar{\tau}(x_n, c))_{ii} = \epsilon \sgn({\rm{diag}}(\bar{\Pi}_{c} - \alpha_{{\rm {med}}}I)_i).
\end{align*}
It can be verified that this choice of $\bar{\tau}(x_n, c)$ yields a physical density matrix $\lambda$. As before, the condition $\epsilon \leq \Delta$ ensures that the minimum eigenvalue of $\lambda$ is positive.  

We now proceed with the same steps as in the previous proof for the $D_1(\cdot, \cdot)$ attack.  We define $
    Q_{\boldsymbol{\sigma}, c} = \sum_{n=1}^T \sigma_n \mathds{1}(c_n = c) \bar{\tau}(x_n,c)$ and use it to re-write $\Delta \mathcal{R}_{p,\epsilon}(\mathcal{M})$. Applying H{\"o}lder's inequality and evaluating the trace norm $\Vert Q_{\boldsymbol{\sigma},c}\Vert_1$ as the sum of the absolute values of its diagonal elements, we arrive at an inequality analogous to (\ref{eq:proof_13}), namely
\begin{align}
    \Delta\mathcal{R}_{\infty,\epsilon}(\mathcal{M}) \leq \frac{d}{T}\Ebb_{\mathcal{T} }\Ebb_{\boldsymbol{\sigma}}\biggl[ \sum_c  \Bigl | \sum_{n=1}^T \epsilon\sigma_n \mathds{1}(c_n = c) \Bigr | \biggr]. 
\end{align}
Again, following the same steps as before, we get
\begin{gather*}
    \Delta\mathcal{R}_{\infty,\epsilon}(\mathcal{M}) \leq d\epsilon\sqrt{\frac{K}{T}}.
\end{gather*}
Using this in \eqref{eq:decomposition}, together with the upper bound in \eqref{eq:banchi_result} returns the upper bound of \eqref{eq:main_result2}.

\section{Proof of Theorem~\ref{thm:generalresult}}\label{app:B}
To obtain the required bound, we proceed as in the proof of Theorem~\ref{thm:1} in Appendix~\ref{app:theorems_12}.  An  upper bound on the adversarial Rademacher complextiy $\mathcal{R}_{p,\epsilon}(\mathcal{M})$ can be obtained as in \eqref{eq:decomposition},  in terms of the standard Rademacher complexity and the perturbation Rademacher complexity. The latter then evaluates as in \eqref{eq:proof_12}. Let $\tau^*(x_n,c)$ denote the perturbation matrix that achieves the inner maximization in \eqref{eq:proof_12}. Subsequently, defining $
    Q_{\boldsymbol{\sigma}, c} = \sum_{n=1}^T \sigma_n \mathds{1}(c_n = c) \tau^{*}(x_n,c),
$ we re-write $\Delta\mathcal{R}_{p, \epsilon}(\mathcal{M}) $ as
\begin{align*}
       \Delta\mathcal{R}_{p, \epsilon}(\mathcal{M}) = {\mathbb{E}}_{\mathcal{T}}{\mathbb{E}}_{\boldsymbol{\sigma}} \biggl[{\sup}_{\Pi \in \mathcal{M}} \frac{1}{T} \sum_c \rm Tr ( \Pi_c Q_{c,\boldsymbol{\sigma}})\biggr].
\end{align*}
Employing H{\"o}lder's inequality yields that ${\rm Tr} ( \Pi_c Q_{c,\boldsymbol{\sigma}}) \leq \Vert \Pi_c\Vert_2 \Vert Q_{c,\boldsymbol{\sigma}} \Vert_2 \leq \sqrt{d} \Vert Q_{c,\boldsymbol{\sigma}} \Vert_2 $, where the last inequality follows since $0 \leq \Pi_c \leq I$. This results in the following upper bound
\begin{align}
    \Delta\mathcal{R}_{p, \epsilon}(\mathcal{M}) \leq {\mathbb{E}}_{\mathcal{T}} {\mathbb{E}}_{\boldsymbol{\sigma}}\Bigl[{\sup}_{\Pi \in \mathcal{M}} \frac{1}{T} \sum_c  \sqrt{d} \lVert Q_{c,\sigma} \rVert_2\Bigr]. \label{eq:proof_2_1}
\end{align}
We now evaluate the 2-norm $\lVert Q_{c,\sigma} \rVert_2$, which can be written as
\begin{align}
    &\lVert Q_{c,\boldsymbol{\sigma}} \rVert_2 \nonumber \\& = \biggl({\rm Tr} \biggl(\sum_{n, m = 1}^T \sigma_n\sigma_m \mathds{1}(c_m = c_n = c)\tau^*(x_n, c)\tau^*(x_m, c)\biggr)\biggr)^{\frac{1}{2}} \nonumber \\&
 =\Bigg( {\rm Tr} \biggl(\sum_{n=1}^T \mathds{1}(c_n = c)\tau^*(x_n, c)^2  + \sum_{n \neq m} (\mathds{1}(\sigma_n\sigma_m = 1) \nonumber \\&- \mathds{1}(\sigma_n\sigma_m = -1)) (\mathds{1}(c_m = c_n = c)\tau^*(x_n, c)\tau^*(x_m, c)) \biggr)\Bigg)^\frac12 \label{eq: Q^2}.
\end{align}
In the following subsections, we consider the two cases $p=1$ and $p=\infty$, and obtain respective upper bounds on \eqref{eq: Q^2}.

Furthermore, using the shorthand notation $\tau_n = \tau^{*}(x_n, c)$, we note that
\begin{gather}
    - \lVert\tau_n \tau_m\rVert_1 \leq \rm Tr (\tau_n \tau_m) \leq\lVert\tau_n \tau_m\rVert_1 \label{eq:interm}
\end{gather}
which will be used to obtain a worst case upper bound on $\lVert Q_{c,\boldsymbol{\sigma}} \rVert_2$. 
\subsection{$D_{1}(\cdot, \cdot)$ perturbation Rademacher complexity}
Using H{\"o}lder's inequality, we get that
$
    \lVert \tau_n \tau_m\rVert_1 \leq \lVert \tau_n \rVert_1 \lVert \tau_m \rVert_\infty 
$, where $\lVert \tau_n \rVert_1 \leq \epsilon$. We now consider $\lVert \tau_m \rVert_\infty$. We write $\tau_m$ in its diagonal basis via a unitary transform $U_m$ as a matrix of positive (P) and a matrix of negative (N) eigenvalues $\tau_m = U_m(P+N)U_m^\dagger$. The trace condition $\rm Tr \tau_m = 0$ implies that $\Vert P \Vert_1 = \Vert N \Vert_1 \leq \epsilon/2$. The $\infty$-Schatten norm gives the eigenvalue of $\tau_m$ with maximal absolute value, which according to the norm bounds on $P$ and $N$ cannot exceed $\epsilon/2$. Thus $\Vert \tau_m \Vert_\infty \leq \epsilon/2$ which together with $\Vert \tau_m \vert_1 \leq \epsilon$ yields $ \lVert \tau_n \tau_m\rVert_1 \leq \epsilon^2/2$. Using this in \eqref{eq:interm} yields
\begin{gather*}
    -\frac{\epsilon^2}{2} \leq {\rm Tr}( \tau_n \tau_m) \leq \frac{\epsilon^2}{2} \; \forall \; n,m.
\end{gather*}
Using this in \eqref{eq: Q^2}, we get the following worst case upper bound:
\begin{align*}
    \lVert Q_{c,\boldsymbol{\sigma}} \rVert_2^2 \leq \frac{\epsilon^2}{2} \Bigg(\sum_{n=1}^T \mathds{1}(c_n = c)  + \sum_{n \neq m} \mathds{1}(c_m = c_n = c) \Bigg).
\end{align*}
Plugging this in \eqref{eq:proof_2_1} for $p=1$, and assuming $K$ equiprobable classes,  we can now upper bound $\mathcal{R}_{1, \epsilon}(\mathcal{M}) $ as
\begin{gather*}
    \mathcal{R}_{1, \epsilon}(\mathcal{M}) \leq \frac{\sqrt{d}K}{T} \mathbb{E}_{\mathcal{T}} {\mathbb{E}}_{\boldsymbol{\sigma}}\mathbb{E}_c \\ \biggl[  \left(\frac{\epsilon^2}{2}\Bigg(\sum_{n=1}^T \mathds{1}(c_n = c) + \sum_{n \neq m} \mathds{1}(c_m = c_n = c) \Bigg)\right)^\frac{1}{2} \biggr].
\end{gather*}
Finally, taking the expectation over the training set $\mathcal{T}$ inside the square root by application of Jensen's inequality yields the following upper bound
\begin{align*}
    \mathcal{R}_{1, \epsilon}(\mathcal{M}) \leq \sqrt{\frac{\epsilon^2}{2} d(1 + \frac{K-1}{T})}.
\end{align*}
Plugging this in \eqref{eq:decomposition} and using the upper bound in \eqref{eq:banchi_result} yields the required bound.
\subsection{$D_{\infty}(\cdot, \cdot)$ perturbation Rademacher complexity}
Under the $D_{\infty}(\cdot, \cdot)$ distance we have that
$
    \lVert\tau_n\rVert_\infty \leq \epsilon$, which implies that $\lVert\tau_n\rVert_1 \leq \epsilon d $. 
Using H{\"o}lder's inequality, we have
$
    \lVert \tau_n \tau_m\rVert_1 \leq \lVert \tau_n \rVert_1 \lVert \tau_m \rVert_\infty, 
$
which together with \eqref{eq:interm} yields:
\begin{gather*}
    -\epsilon^2d \leq {\rm Tr}( \tau_n \tau_m) \leq \epsilon^2d \; \forall \; n,m.
\end{gather*}
Using this, we get the following worst case upper bound on $\lVert Q_{c,\boldsymbol{\sigma}} \rVert_2^2$
\begin{align*}
    \lVert Q_{c,\boldsymbol{\sigma}} \rVert_2^2 \leq \epsilon^2 d \Bigg(\sum_{n=1}^T \mathds{1}(c_n = c)  + \sum_{n \neq m} \mathds{1}(c_m = c_n = c) \Bigg).
\end{align*}

Retracing the remaining steps of the $D_{1}(\cdot, \cdot)$ adversary derivation yields
\begin{align*}
    \mathcal{R}_{\infty, \epsilon}(\mathcal{M}) \leq \epsilon d\sqrt{1 + \frac{K-1}{T}}
\end{align*}
which together with \eqref{eq:decomposition} and \eqref{eq:banchi_result} concludes the proof.
\section{Proof of Lemma~\ref{lem:strength}}
To derive the required relation, we start by noting that $p$-Schatten distance between the states $\rho$ and $\lambda$ is defined as $D_p(\rho, \lambda) = \Vert \rho-\lambda\Vert_p.$  
Thus, defining the matrix $\tau = \rho-\lambda$, the distance condition can be written as $$D_p(\rho,\lambda)=\Vert \tau \Vert_p \leq \epsilon.$$ Furthermore, we remind ourselves of the generalized version of H{\"o}lder's inequality \cite{trace_ineq} for matrices $A, \; B$ $$\Vert AB\Vert_r \leq \Vert A\Vert_p\Vert B\Vert_q, \; 1/r=1/p+1/q.$$

Firstly we prove the inequality for the case $p\leq p'$. Assume that $D_p(\rho,\lambda) \leq \epsilon$ for some $\epsilon\geq 0$.  Then the following set of relations hold,
\begin{align}
    \Vert \tau\Vert_{p'}  = \Vert I\;\tau\Vert_{p'}  &\leq \Vert \tau\Vert_p'\Vert I\Vert_{pp'/(p'-p)} \nonumber \\& = d^{1/p-1/p'}\Vert \tau\Vert_p\nonumber, \end{align}where in the first line we inserted the identity operator $I$ and in the second applied H{\"o}lder's inequality.  This in turn implies that  \begin{align}D_{p}(\rho, \lambda) \geq d^{1/p'-1/p} D_{p'}(\rho, \lambda) \label{eq:D_rel}
\end{align}

The inequality \eqref{eq:D_rel} implies that a $p$-adversary with perturbation budget $\epsilon$ can access all states within the set $$\{\lambda:D_{p'}(\rho, \lambda) \leq d^{1/p'-1/p}\epsilon\}.$$
Thus, if we have a $p$-adversary with perturbation budget $\epsilon$ at training, and a $p'$-adversary with perturbation budget $\epsilon'$ during testing, the training adversary is stronger than the testing adversary if 
\begin{align}
    \epsilon' < d^{1/p'-1/p}\epsilon.
\end{align}

We now proceed to the case of $p>p'$. Consider the set $\{\lambda: D_p(\lambda, \rho)\leq \epsilon\}$. With the definitions above, this can be equivalently written as $\lVert\tau\rVert_p \leq \epsilon$. Holder's inequality implies that $\lVert\tau\rVert_1 \leq \lVert\tau\rVert_p d^{1-1/p}$. Furthermore, as shown in appendix \ref{app:B}, $\lVert\tau_\infty \leq \lVert\tau\rVert_1/2$. Thus $$\lVert\tau\rVert_p \leq\epsilon \implies \lVert\tau\rVert_\infty \leq \frac{\epsilon}{2}d^{(1-1/p)}.$$ To lower bound $\lVert\tau\rVert_\infty$ we again use H{\"o}lder's inequality; $\lVert\tau\rVert_p \leq\lVert\tau\rVert_\infty\lVert I\rVert_p$. Thus $$\lVert\tau\rVert_p \leq \epsilon \implies \lVert\tau\rVert_\infty\geq\epsilon d^{(-1/p)}.$$ Therefore a sufficient condition to enforce the inclusion condition is that the lower bound on the $\infty$-Schatten norm of the $p$-adversarial attack is greater than the upper bound on the $\infty$-Schatten norm of the $p'$-adversarial attack
\begin{align}
    \epsilon'<2d^{(1/p'-1/p-1)}\epsilon
\end{align}
This concludes the proof.
\section{Proof of Theorem~\ref{thm:mismatched}}
To obtain bounds on the mismatched adversarial generalization error, we first decompose it as
 \begin{align}  \mathcal{G}_{p,p',\epsilon,\epsilon'}(\Pi,\mathcal{T}) =  \nu_{p,p';\epsilon,\epsilon'}(\Pi)+ \mathcal{G}_{p,\epsilon}(\Pi,\mathcal{T}), \label{eq:decomposition_mismatched}
\end{align} where $\nu_{p,p';\epsilon,\epsilon'}(\Pi)=L_{p',\epsilon'}(\Pi)-L_{p,\epsilon}(\Pi)$ is the difference in adversarial population risks due to adversarial mismatch and $ \mathcal{G}_{p,\epsilon}(\Pi,\mathcal{T})$, defined as in \eqref{eq:adversarial_generror}, is the adversarial generalization error with no adversarial mismatch. 

We now derive the bounds stated in Theorem~\ref{thm:mismatched}. To this end we consider the following two cases:
\begin{itemize}
    \item The training adversary is stronger than the testing adversary, as defined in Definition \ref{def:strenght}, i.e. $\{\lambda:D_p(\lambda, \rho(x))\leq\epsilon\} \supset \{\lambda:D_{p'}(\lambda, \rho(x))\leq\epsilon'\}$. Then we have 
    \begin{gather*}          \underset{P(x,c)}{\Ebb}\Bigl[\underset{\lambda:D_p(\lambda, \rho(x))\leq \epsilon}{\min}({\rm Tr}\Pi_c \lambda)\Bigr]\\ \leq \underset{P(x,c)}{\Ebb}\Bigl[\underset{\lambda':D_{p'}(\lambda', \rho(x))\leq \epsilon'}{\min}({\rm Tr}\Pi_c \lambda') \Bigr].
    \end{gather*} This  implies that $\nu_{p, p';\epsilon, \epsilon'}(\Pi) \leq 0$. Using this in \eqref{eq:decomposition_mismatched} yields the following upper bound $\mathcal{G}_{p,p',\epsilon,\epsilon'}(\Pi,\mathcal{T}) \leq \mathcal{G}_{p,\epsilon}(\Pi,\mathcal{T})$.
    \item The training adversary is weaker than the testing adversary i.e. $\{\lambda:D_p(\lambda, \rho(x))\leq\epsilon\} \subset \{\lambda:D_{p'}(\lambda, \rho(x))\leq\epsilon'\}$. Then we have 
    \begin{gather*}          \underset{P(x,c)}{\Ebb}\Bigl[\underset{\lambda:D_p(\lambda, \rho(x))\leq \epsilon}{\min}({\rm Tr}\Pi_c \lambda)\Bigr]\\ \geq \underset{P(x,c)}{\Ebb}\Bigl[\underset{\lambda':D_{p'}(\lambda', \rho(x))\leq \epsilon'}{\min}({\rm Tr}\Pi_c \lambda') \Bigr]
    \end{gather*}
    This in turn implies that $\nu_{p, p';\epsilon, \epsilon'}(\Pi) \geq 0.$
    Combining this with \eqref{eq:decomposition_mismatched} gives the following lower bound on the mismatched adversarial generalization error $\mathcal{G}_{p,p',\epsilon,\epsilon'}(\Pi,\mathcal{T}) \geq \mathcal{G}_{p,\epsilon}(\Pi,\mathcal{T})$ in the case of a weak training adversary
\end{itemize}

To obtain a lower bound in the case of a strong training adversary, and an upper bound in the case of a weak adversary, we need to bound the term $\nu_{p, p';\epsilon,\epsilon'}(\Pi)$. To this end, we start by expressing the mismatch as
\begin{align}
    \nu_{p,p';\epsilon,\epsilon'}(\Pi) &= \Ebb_{P(x,c)}[\ell_{p',\epsilon'}(\Pi,\rho(x),c) - \ell_{p,\epsilon}(\Pi,\rho(x),c)]  \nonumber \\
    & = \Ebb_{P(x,c)}\Bigl[ {\rm Tr}\Bigl(\Pi_c (\lambda^*_{p,\epsilon}(x,c)-\lambda^*_{p',\epsilon'}(x,c)) \Bigr) \Bigr]\nonumber \\ & = \Ebb_{P(x,c)} \Bigl[{\rm Tr}(\Pi_c \Delta \lambda_{p, p';\epsilon,\epsilon'}(x, c)) \Bigr]\nonumber = C, 
\end{align} where $\lambda^*_{p,\epsilon}(x,c)$ is the optimal adversarial example for a $p$-adversarial attack with perturbation budget $\epsilon$, and likewise for $\lambda^*_{p',\epsilon'}(x,c)$. The matrix $\Delta \lambda_{p, p';\epsilon,\epsilon'}(x, c)$ is a trace zero matrix thus $C$ may take both positive and negative values. Using this, we write
\begin{align} \label{eq:1} -|C| \leq \nu_{p,p',\epsilon,\epsilon'}(\Pi)\leq |C|.\end{align}
We now proceed to upper bound the term $C$.
\begin{align}
&|C| = |\Ebb_{P(x,c)}[ {\rm Tr}(\Pi_c \Delta \lambda_{p, p';\epsilon,\epsilon'}(x, c))]| \nonumber\\
& \leq \Ebb_{P(x,c)}[|{\rm Tr}(\Pi_c \Delta \lambda_{p, p';\epsilon,\epsilon'}(x, c))|]\nonumber\\
&\leq\Ebb_{P(x,c)}[\Vert \lambda^*_{p,\epsilon}(x,c)-\lambda^*_{p',\epsilon'}(x,c)\Vert_1],
\end{align}
where the first inequality follows from Jensen's inequality, and the second inequality follows from H{\"o}lder's using $\Vert \Pi_c \Vert_\infty \leq 1$.
The trace norm $\Vert \lambda^*_{p,\epsilon}(x,c)-\lambda^*_{p',\epsilon'}(x,c)\Vert_1$ can be further upper bounded as
\begin{align}
&\Vert \lambda^*_{p,\epsilon}(x,c)-\lambda^*_{p',\epsilon'}(x,c)\Vert_1 \nonumber \\& \leq  \Vert \lambda^*_{p,\epsilon}(x,c)- \rho(x) \Vert_1 +\Vert \rho(x)-\lambda^*_{p',\epsilon'}(x,c)\Vert_1 \nonumber\\
& \leq d^{(1-1/p)}\Vert \lambda^*_{p,\epsilon}(x,c)- \rho(x) \Vert_p \nonumber \\& + d^{(1-1/p')}\Vert \lambda^*_{p',\epsilon'}(x,c)- \rho(x) \Vert_{p'} \nonumber \\
& \leq d^{(1-1/p)}\epsilon+ d^{(1-1/p')}\epsilon' = \xi,
\end{align} 
where the first inequality follows by adding and subtracting the term $\rho(x)$ and then applying the triangle inequality. The second inequality is an application of H{\"o}lder's inequality with  $\Vert I \Vert_{p/(p-1)} \leq d^{1-1/p}$. The last inequality follows since $\lambda^*_{p,\epsilon}(x,c)$ is the optimal perturbed quantum state satisfying the constraint,  $\Vert \lambda^*_{p,\epsilon}(x,c)- \rho(x) \Vert_p \leq \epsilon.$

Thus we can bound the mismatch $\nu_{p,p',\epsilon,\epsilon'}(\Pi)$ as follows: $$-\xi\leq\nu_{p,p',\epsilon,\epsilon'}(\Pi)\leq \xi.$$ 
Using this again in \eqref{eq:decomposition_mismatched} gives the following lower bound for when the training adversary is stronger,
$$ \mathcal{G}_{p,\epsilon}(\Pi,\mathcal{T}) -\xi \leq \mathcal{G}_{p,p',\epsilon,\epsilon'}(\Pi,\mathcal{T}),$$ and the following upper bound for when the training adversary is weaker,
$$ \mathcal{G}_{p,p',\epsilon,\epsilon'}(\Pi,\mathcal{T}) \leq \mathcal{G}_{p,\epsilon}(\Pi,\mathcal{T}) +\xi.$$
This concludes the proof.

\section{Noisy Quantum Embedding Satisfies Assumption~\ref{asu1}}
\label{noise_asu1_apdx}
In this section, we show that the minimum eigenvalue of the quantum state $\rho'(x)=\mathcal{E}(\rho(x))$ resulting due to a noisy quantum embedding $x \mapsto \mathcal{E}(\rho(x))$ is at least the minimum eigenvalue of the noiseless state $\rho(x)$. To this end,  we note that the CPTP map $\mathcal{E}(\cdot)$ can be equivalently written as $\mathcal{E}(\rho)=\sum_{i}E_i \rho E_i^{\dag}$ where $\{K_i\}$ is the set of Kraus operators satisfsying the completeness relation, $\sum_{i}K_i^{\dag}K_i =I$.

To compute 
the minimal eigenvalue of the noisy state $\mathcal{E}(\rho)$, we use the variational principle as
\begin{align}
    \alpha_{\min}(\mathcal{E}(\rho)) &= \underset{|\psi\rangle}{\min}\sum_i \frac{\langle\psi|E_i \rho E_i^\dagger|\psi\rangle}{\langle\psi|\psi\rangle} \nonumber \\
 &=\underset{|\psi\rangle}{\min}\sum_i \frac{\langle\psi|E_i \rho E_i^\dagger|\psi\rangle}{ \langle\psi|E_iE_i^\dagger|\psi\rangle}\frac{\langle\psi|E_iE_i^\dagger|\psi\rangle}{\langle\psi|\psi\rangle}. \label{eq:proof_a}
\end{align}
In \eqref{eq:proof_a}, the first term is an upper bound on the minimal eigenvalue of $\rho$. Using this, we have
\begin{gather*}
    \alpha_{\min}(\mathcal{E}(\rho)) \geq \sum_i \alpha_{\min}(\rho)\frac{\langle\psi|E_iE_i^\dagger|\psi\rangle}{\langle\psi|\psi\rangle}= \alpha_{\min}(\rho).
\end{gather*}
Furthermore, equality holds only if there exists a state $\vert\alpha^{\min}\rangle$ such that 
\begin{align*}
    \vert \alpha^{\min}\rangle = \arg \min_{\vert \psi\rangle} \frac{\langle \psi \vert E_i\rho E^\dagger_i\vert\psi\rangle}{\langle \psi \vert E_i E_i^\dagger \vert\psi\rangle}\; \forall \; i. 
\end{align*}
This is because in equation (\ref{eq:proof_a}) if $\vert\alpha^{\min}\rangle$ does not exist, then
\begin{align*}
    \frac{\langle \psi \vert E_i\rho E^\dagger_i\vert\psi\rangle}{\langle \psi \vert E_i E_i^\dagger \vert\psi\rangle} = \alpha_{\min}(\rho)
\end{align*}
is not attainable for all $i$ simultaneously, hence $\alpha_{\min}(\mathcal{E}(\rho))=\alpha_{\min}(\rho) $ cannot be achieved.

\end{document}